\documentclass[aps,prl,twocolumn,superscriptaddress]{revtex4}
\usepackage{graphicx}
\usepackage{amssymb}
\usepackage{amsmath}
\begin{document}

\def\bc{\begin{center}}
\def\ec{\end{center}}
\def\bi{\begin{itemize}}
\def\ei{\end{itemize}}
\def\bd{\begin{description}}
\def\ed{\end{description}}
\def\noi{\noindent}
\def\ba{\begin{array}}
\def\ea{\end{array}}
\preprint{}
\title{ Thermal quantum  and classical correlations in two qubit XX model in a nonuniform external magnetic field.}

\author{Ali Saif M. Hassan}
\email{alisaif73@gmail.com}
 \affiliation{Department of Physics, University of Amran, Amran, Yemen}
\author{Behzad Lari}
\email{behzadlari1979@yahoo.com}
%\affiliation{Department of Physics, University of Pune, Pune, India-411007}

\author{Pramod S. Joag}
\email{pramod@physics.unipune.ac.in}
\affiliation{Department of Physics, University of Pune, Pune, India-411007.}

\date{\today}
\begin{abstract}

 We investigate how thermal quantum discord $(QD)$ and classical correlations $(CC)$ of  a two qubit one-dimensional XX Heisenberg chain in thermal equilibrium depend on temperature of the bath as well as on nonuniform external magnetic fields applied to two qubits and varied separately. We show that the behaviour of $QD$ differs in many unexpected ways from thermal entanglement $(EOF)$. For the nonuniform case, $(B_1= - B_2)$ we find that $QD$ and $CC$ are equal for all values of $(B_1=-B_2)$ and for different temperatures. We show that, in this case, the thermal states of the system belong to a class of mixed states and satisfy certain conditions under which $QD$ and $CC$ are equal. The specification of this class and the corresponding conditions are completely general and apply to any quantum system in a state in this class and satisfying these conditions. We further find that the relative contributions of $QD$ and $CC$ can be controlled easily by changing the relative magnitudes of $B_1$ and $B_2$. Finally, we connect our results with the monogamy relations between the EOF, classical correlations and the quantum discord of two qubits and the environment. \\

\noi PACS numbers: 03.65.Ud;75.10.Pq;05.30.-d
 \end{abstract}

\maketitle

%\emph{Introduction:}

It is now quite well known that composite quantum systems can be in a class of states, called entangled states, in which the correlations between the constituents of the system cannot be achieved in a classical world \cite{nie,wer}. Although all pure entangled states possess such nonlocal quantum correlations, there are mixed entangled states which do not, in the sense of violating Bell inequalities \cite{bell}. The entanglement in quantum states and the resulting nonlocal quantum correlations form an area of intense research, due to their huge technological promise, especially in the areas of quantum communication and cryptography \cite{eker}. However, quantum correlations breaking Bell inequalities need not account for all quantum correlations in a composite quantum system in a given state. In order to account for the quantum correlation in a given state, we must find some means to divide the total correlation into a classical part and a purely quantum part. This is particularly important for mixed states, since their quantum correlations are many a time hidden by their classical correlations $(CC)$. An answer to this requirement is given by quantum discord $(QD)$, a measure of the quantumness of correlations introduced in Ref. \cite{olli}. Quantum discord is built on the fact that two classically equivalent ways of defining the mutual information turn out to be inequivalent in the quantum domain. In addition to its conceptual role, some recent results \cite{knill}, suggest that quantum discord and not entanglement may be responsible for the efficiency of a mixed state based quantum computer. Present authors believe that $QD$ will turn out to be a very useful tool to analyze mixed state quantum correlations and their consequences, as  mixed state entanglement is very difficult and eluding to deal with \cite{hass}. To  realize this hope we need a viable relation between mixed state entanglement and quantum discord \cite{modi}. The pointers towards such a relation may be obtained by studying these properties for various quantum systems.

Motivated by these considerations, we present here the results of our investigation of the amount of $QD$ and $CC$ in a two qubit Heisenberg XX chain at finite temperature subjected to nonuniform external magnetic fields $B_1$ and $B_2$ acting separately on each qubit. We study two distinct cases namely $B_1=-B_2$ (nonuniform field) and $B_1=B_2$ (uniform field). In each case, we obtain the dependance of $QD, CC$ and entanglement of formation $(EOF)$ \cite{ben} in the system on the external magnetic field and temperature. Such a model is realized, for example, by a pair of qubits (spin 1/2) within a solid at finite temperature experiencing a spatially varying magnetic field. Such Heisenberg models can describe fairly well the magnetic properties of real solids \cite{hamm} and are well adapted to the study of the interplay of disorder and entanglement as well as of entanglement and quantum phase transitions \cite{sach,sar}. The variation of entanglement \cite{sun} and $QD$ \cite{werl,maz} in a two qubit Heisnberg XX chain with external magnetic field is already reported.

In order to quantify entanglement in a thermally mixed two qubit state, we use entanglement of formation derived using concurrence given by \cite{hill}
 $$C=\max\{\lambda_1-\lambda_2-\lambda_3-\lambda_4,0\} \eqno{(1)}$$
where $\lambda_i (i=1,2,3,4)$ are the square roots of the eigenvalues
of the operator $\rho \tilde{\rho}$ in descending order $$\tilde{\rho}=(\sigma_1^y\otimes \sigma_2^y)\rho^{\ast}(\sigma_1^y\otimes \sigma_2^y),\eqno{(2)}$$ with $ \lambda_1\geq \lambda_2 \geq \lambda_3 \geq \lambda_4,$ and $\rho$ is the density matrix of the pair qubits; $\sigma_1^y$ and $\sigma_2^y$ are the normal Pauli operators. The entanglement of formation is related to the concurrence by $$EN=h\left(\frac{1+\sqrt{1-C^2}}{2}\right),$$ where $h(x)=-x\log_2 x-(1-x)\log_2 (1-x).$ Henceforth, in this paper, we denote the entanglement of formation ($EOF$) by $EN.$ The concurrence $C=0$ corresponds to an unentangled state and $C=1$ corresponds to a maximally entangled state. \\

\noi \emph{The thermalized Heisenberg system.}
The model Hamiltonian we study is given by $$H=J(S_1^xS_2^x+S_1^yS_2^y)+B_1S_1^z+B_2S_2^z, \eqno{(3)}$$
where $S^{\alpha}\equiv\sigma^{\alpha}/2,~~~(\alpha=x,y,z)$ are the spin $1/2$ operators, $\sigma^{\alpha}$ are the Pauli operators and $J$ is the strength of Heisenberg interaction. $B_1$ and $B_2$ are external magnetic fields. As stated in the introduction, by changing $B_1$ and $B_2$ separately, we want to study the effects of magnetic field on the thermal $QD, CC$ and $EN$ in a very general way. The eigenvalues and eigenvectors of H are

$$H |00\rangle = -(B_1+B_2)|00\rangle,$$
$$H |11\rangle= (B_1+B_2)|11\rangle,$$
$$ H |\psi^{\pm}\rangle=\pm D |\psi^{\pm}\rangle,\eqno{(4)}
 $$
 where $D^2= (B_1-B_2)^2+J^2$
 and $|\psi^{\pm}\rangle=\frac{1}{N_{\pm}}[|01\rangle+\frac{(B_1-B_2)\pm D}{J} |10\rangle].$
 We denote the eigenvalues corresponding to $|00\rangle, |11\rangle,|\psi^{\pm}\rangle$ by $E_{00},E_{11},E_{\pm}$ respectively. In the standard basis, $\{ |00\rangle, |01\rangle, |10\rangle, |11\rangle\},$ the density matrix $\rho(T)$ is given by
 $$ \rho(T)=\frac{1}{Z} \left[ \ba{rrrr} u_1 & 0 & 0 & 0 \\ 0 & w_1 & v & 0 \\ 0 & v & w_2 & 0 \\ 0 & 0 & 0 & u_2 \ea \right], \eqno{(5)}$$

 where
 $$u_1=e^{(B_1+B_2)/{kT}},$$
 $$u_2=e^{-(B_1+B_2)/{kT}},$$
$$w_1=\cosh(\frac{D}{kT})+\frac{(B_1-B_2)}{D}\sinh(\frac{D}{kT}),$$
$$w_2=\cosh(\frac{D}{kT})-\frac{(B_1-B_2)}{D}\sinh(\frac{D}{kT}),$$
$$v=-\frac{J\sinh(\frac{D}{kT})}{D}, \eqno{(6)}$$
and $Z=Tr[\exp(\frac{-H}{kT})]$ is the partition function. In the following we select $|J|$ as the energy unit and set $k=1.$\\

\noi \emph{Quantum Discord.} \cite{olli,werl} In classical information theory (CIT) the total correlation between two systems (two sets of random variables) A and B described by a joint distribution probability $p(A,B)$ is given by the mutual information (MI), $$I(A, B) = H(A) + H(B) - H(A, B), \eqno{(7)}$$
with the Shannon entropy $H(\cdot) = -\sum_j p_j log_2 p_j$. Here $p_j$ represents the probability of an event $j$ associated to
systems $A, B,$ or to the joint system $AB$. Using Bayes's rule we may write MI as $$I(A, B) = H(A) - H(A|B), \eqno{(8)}$$ where $H(A|B)$ is the classical conditional entropy. In CIT these two expressions are equivalent but in the quantum domain this is no longer true \cite{olli,lan}. The first quantum
extension of MI, denoted by $I (\rho)$, is obtained directly replacing the Shannon entropy in Eq.(7) with the von Neumann entropy, $S (\rho) = -Tr (\rho log_2 \rho)$, with $\rho$, a density matrix, replacing probability distributions. To obtain a quantum version of Eq. (8) it is necessary to generalize the classical conditional entropy. This is done recognizing $H(A|B)$ as a measure of our ignorance about system $A$ after we make a set of measurements on $B$. When $B$ is a quantum system the choice of measurements determines the amount of information we can extract from it. We restrict ourselves to von Neumann measurements on $B$ described by a complete set of orthogonal projectors, $\Pi_j$, corresponding to outcomes $j$.

After a measurement, the quantum state $\rho$ changes to $\rho_j = [(I\otimes \Pi_j)\rho (I\otimes \Pi_j)]/p_j$, with $I$ the identity operator
for system $A$ and $p_j = Tr[(I\otimes \Pi_j)\rho (I\otimes \Pi_j)]$. Thus, one defines the quantum analog of the conditional entropy as $S (\rho|\{\Pi_j\}) = \sum_j p_j S (\rho_j)$ and, consequently, the second quantum extension of the classical MI as \cite{olli}

 $\mathcal{J}(\rho|\{\Pi_j\}) = S(\rho^A)- S(\rho|\{\Pi_j\})$. The value of $\mathcal{J}(\rho|\{\Pi_j\})$ depends on the choice of $\{\Pi_j\}$.

 Henderson and Vedral \cite{olli} have shown that the maximum of $\mathcal{J}(\rho|\{\Pi_j\})$ with respect to $\{\Pi_j\}$ can be interpreted as a measure of classical correlations. Therefore, the difference between the total correlations $I (\rho)$ and the classical correlations
 $\mathcal{Q}(\rho) = sup_{\{\Pi_j\}} \mathcal{J}(\rho|\{\Pi_j\}) $ is defined as $$D(\rho) = I (\rho) - Q(\rho), \eqno{(9)}$$ giving, finally, a measure of quantum correlations \cite{olli} called quantum discord $(QD)$. For pure states $QD$ reduces to entropy of entanglement \cite{ben}, highlighting that in this case all correlations come from entanglement.
However, it is possible to find separable (not-entangled) mixed states with nonzero $QD$ \cite{olli,lan}, meaning that entanglement
does not cause all nonclassical correlations contained in a composite quantum system. Also, $QD$ can be operationally seen as the difference of work that can be extracted from a heat bath using a bipartite system acting either globally or only locally \cite{zur}.

\noi\emph{Results and discussion.}
Case I: $B_1=-B_2,$ and $(J>0)$.

In this case, $|\psi^-\rangle$ is the ground state with eigenvalue $E_-=-\sqrt{4B_1^2+J^2}.$ Other eigenvalues are $\{0, 0, \sqrt{4B_1^2+J^2}\}$ for eigenvectors $\{|00\rangle, |11\rangle, |\psi^+\rangle\}$ respectively.

In this case the variation of $QD, CC$ and  $EN$  with $B_1$ at different  temperatures (T=0.2, 0.9, 1.5) is depicted in Figs. 1, 2, 3, respectively. We observe that $QD$ and $CC$ in the thermal state coincide for all values of $B_1$ for different temperatures (T=0.2, 0.9, 1.5).

In order to understand this observation, we take a close look at the thermal state.  At temperature T, the thermal state is given by

$$\rho=\frac{1}{Z}\big{[}|00\rangle\langle00|+|11\rangle\langle11|+e^{\sqrt{4B_1^2+J^2}/{T}}$$ $$|\psi^-\rangle\langle\psi^-|+e^{-\sqrt{4B_1^2+J^2}/{T}}|\psi^+\rangle\langle\psi^+|\big{]}.\eqno{(10)}$$
This $\rho$ has the Bloch representation \cite{hj}
$$\rho=\frac{1}{4}[I\otimes I+\sum_{i=1}^3 c_i \sigma_i\otimes \sigma_i], \eqno{(11)}$$

where $\sigma_i~~(i=1, 2, 3)$ are the one-qubit Pauli operators. In the appendix to this paper we prove that, the class of mixed states as in Eq.(11) have equal classical and quantum correlations (like bipartite pure states \cite{luo1}) provided $$c_i=c_j>c_k$$ and $$c_k=-c_i^2 \eqno{(12)}$$ where $i\neq j \neq k \in \{1,2,3\}$. Here $c_i, c_j, c_k$ are the diagonal elements of the correlation matrix defined by $c_{ij}=Tr(\rho \sigma_i\otimes \sigma_j).$
It is straightforward to check that the thermal state $\rho$ in Eq.(10) which has form of Eq.(11) satisfies conditions (12).
This explains the observations in Figs. 1, 2, 3, that the two qubit thermal state $\rho$ for $B_1 = -B_2$ in Eq.(10) gives rise to equal $QD$ and $CC$ for all values of $B_1$ and temperature. In order to see why the common curve for $QD$ and $CC$ peaks at $B_1=0$ we can maximize the expression for $QD=CC$  with respect to $B_1$ and check that the maximum occurs at $B_1=0.$

\begin{figure}[!ht]
\begin{center}
\includegraphics[width=8cm,height=5cm]{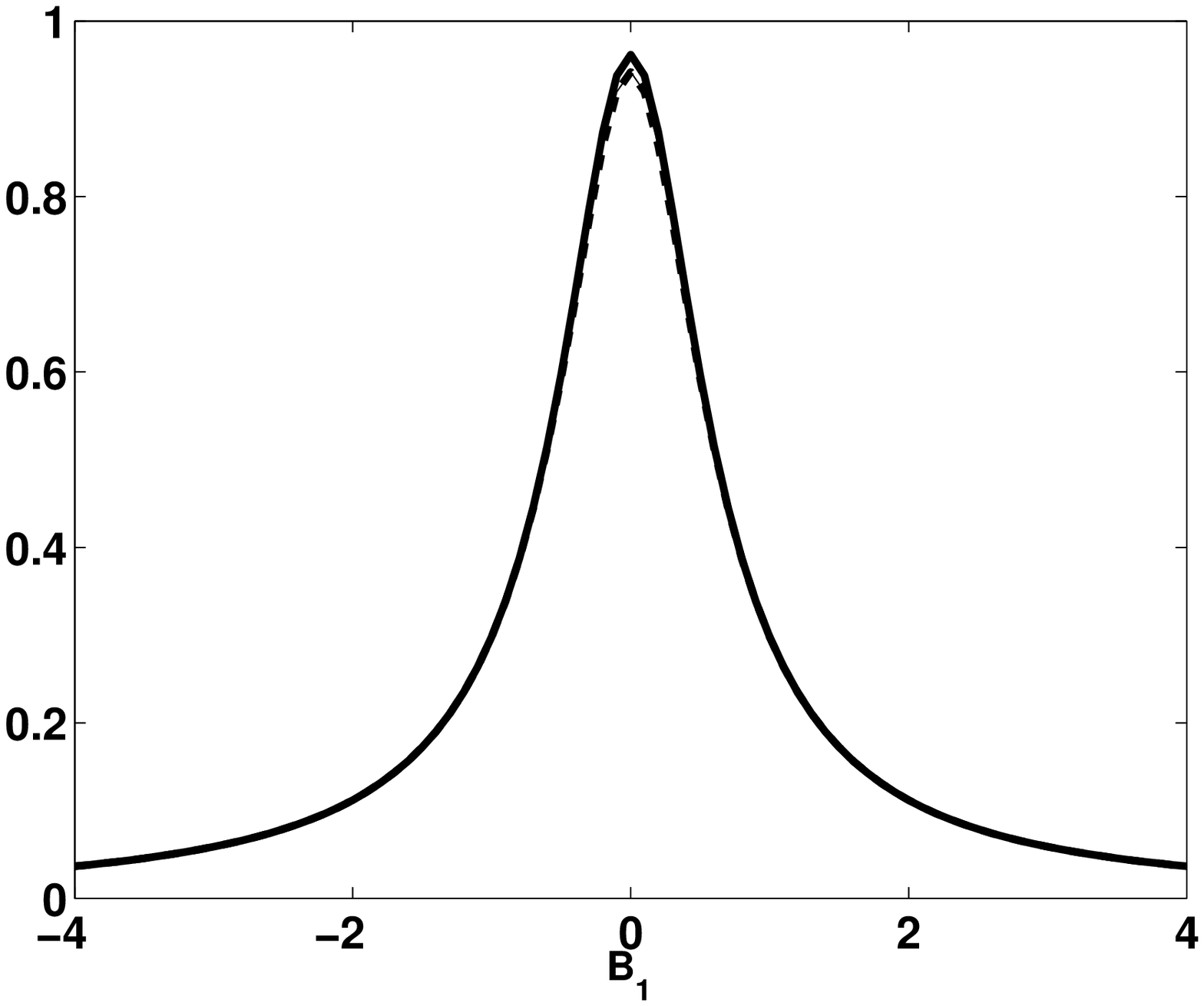}

Figure 1 : $QD$ and $CC$ (dashed line) and $EN$ (solid line) as a function  of external magnetic field $B_1=-B_2$ at $T=0.2$
\end{center}
\end{figure}

\begin{figure}[!ht]
\begin{center}
\includegraphics[width=8cm,height=5cm]{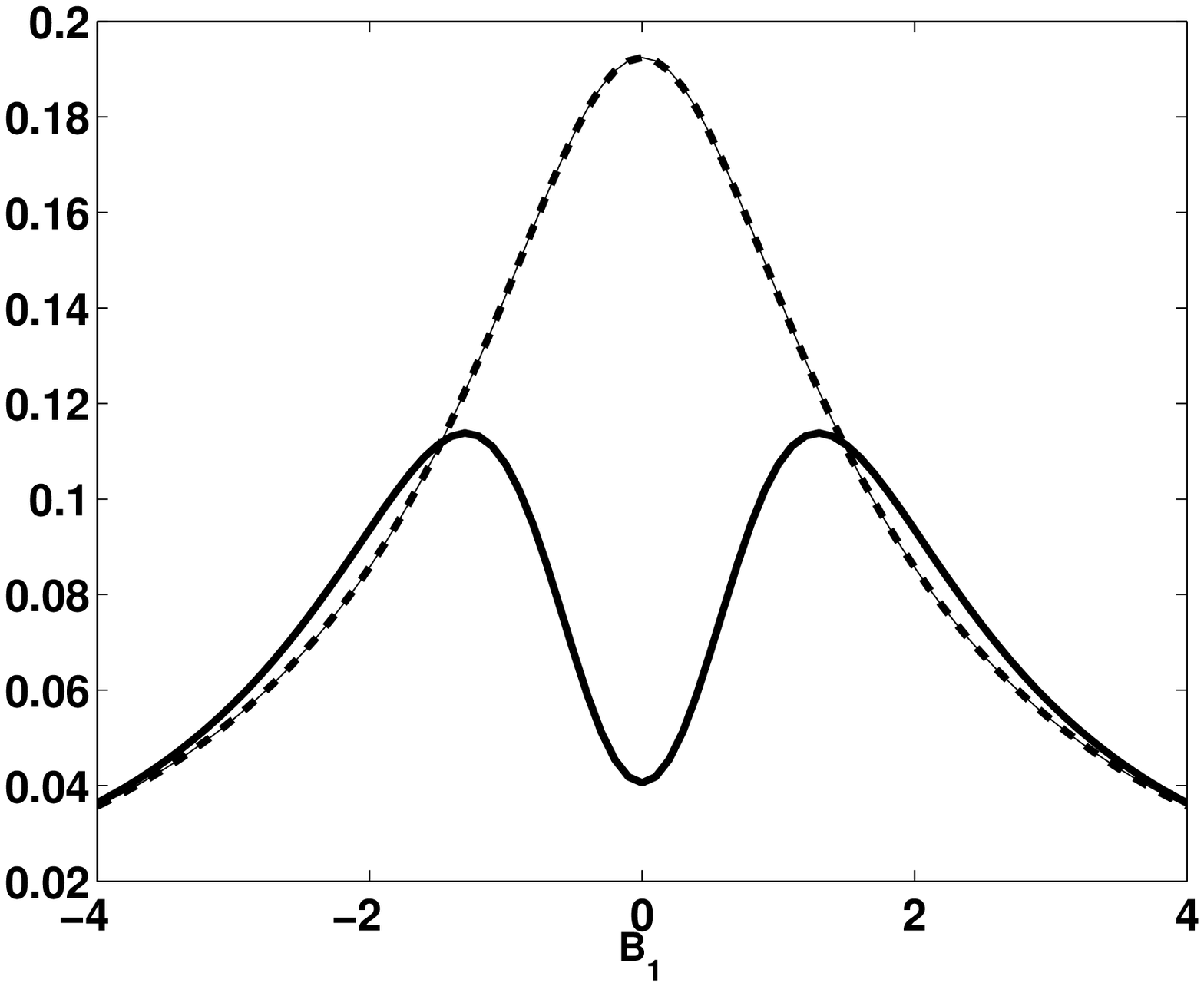}

Figure 2 : $QD$ and $CC$ (dashed line) and $EN$ (solid line) as a function  of external magnetic field $B_1=-B_2$ at $T=0.9$
\end{center}
\end{figure}

\begin{figure}[!ht]
\begin{center}
\includegraphics[width=8cm,height=5cm]{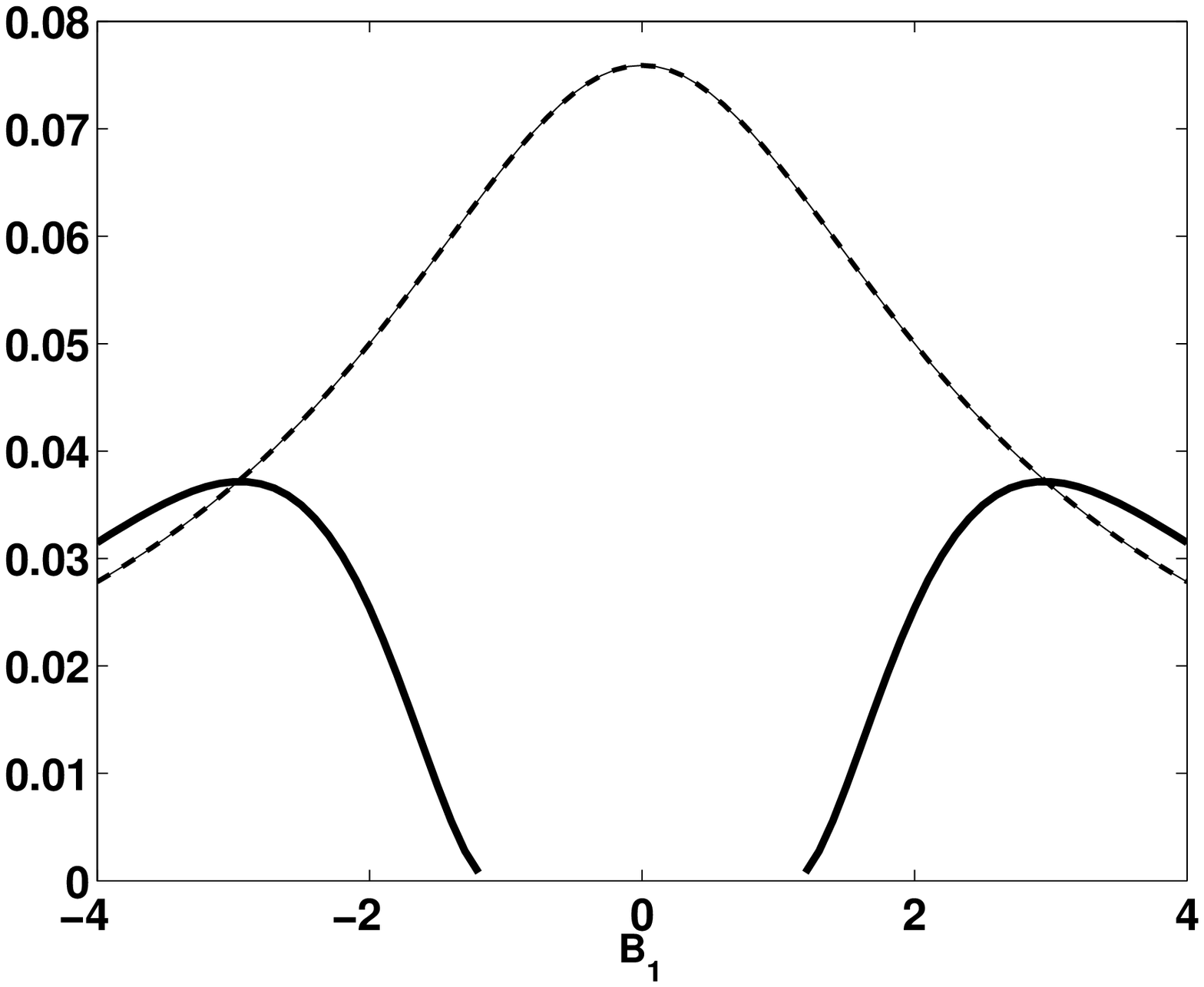}

Figure 3 : $QD$ and $CC$ (dashed line) and $EN$ (solid line) as a function  of external magnetic field $B_1=-B_2$ at $T=1.5$
\end{center}
\end{figure}

  From Figs. 1, 2, 3,  we also see that $EN$ as a function of $B_1$ has a peak at $B_1=0$ for $T=0.2$, has a dip for $T=0.9$ and goes to zero over an interval symmetric about $B_1=0$ for $T=1.5$ \cite{sun}. From Eq. (10) we see that the concurrence of the thermal state is governed by the admixture of the $|\psi^+\rangle$ and  $|\psi^-\rangle$ states. We expect concurrence to fall as the state  $|\psi^+\rangle$ classically mixes more and more with the ground state  $|\psi^-\rangle.$  For fixed $J=1$ and a fixed temperature T, this happens for $B_1=0$. That is why we got a dip in the $EN$ curve at $B_1=0$. The size of this dip increases with temperature. In fact the dip touches the $B_1$ axis when, at $B_1=0$ the concurrence is zero. To see this we note that the concurrence for $\rho$ in Eq.(10) is given by \cite{oconn}
  $$C=\frac{2}{Z}\max\{|v|-\sqrt{u_1 u_2},0\}\eqno{(13)}$$
  where $v, u_1, u_2$ are given in Eq. (6).
  Therefore, for $B_1=0$ and $J=1$, $C\geq0$ provided $\sinh\frac{1}{T} \geq 1$ or $T\leq 1.1346$. For $T=1.5$ and $J=1$, using the requirement $\sinh\frac{D}{T}=D$, we can find the range of $B_1$ around $B_1=0$ in which $C=0$. This is $-1.1456 \leq B_1 \leq 1.1456$. Figs. 1,2,3 confirm the corresponding behaviour of $EN.$

  Fig.4 shows the variation of $EN, QD, CC$  with temperature at fixed values of $B_1=-B_2$. As expected we have $QD=CC$ for all temperatures. Both $EN$ and $QD~ (CC)$ curves have plateau at low temperatures corresponding to their ground state values, as at these temperatures, the ground state is not thermally connected to other exited states. Other interesting observation is  the vanishing of concurrence at a finite critical temperature $T_c$ which increases with $B_1$ value, while $QD$ and $CC$ asymptotically go to zero with temperature. The increase in $T_c$ with $B_1$ \cite{sun} can be understood from the thermal state Eq.(10) which says that higher temperatures are required to get a given admixture of $|\psi^-\rangle$ and $|\psi^+\rangle$ for higher $B_1$ values.

  \begin{figure}[!ht]
\begin{center}
\includegraphics[width=8cm,height=5cm]{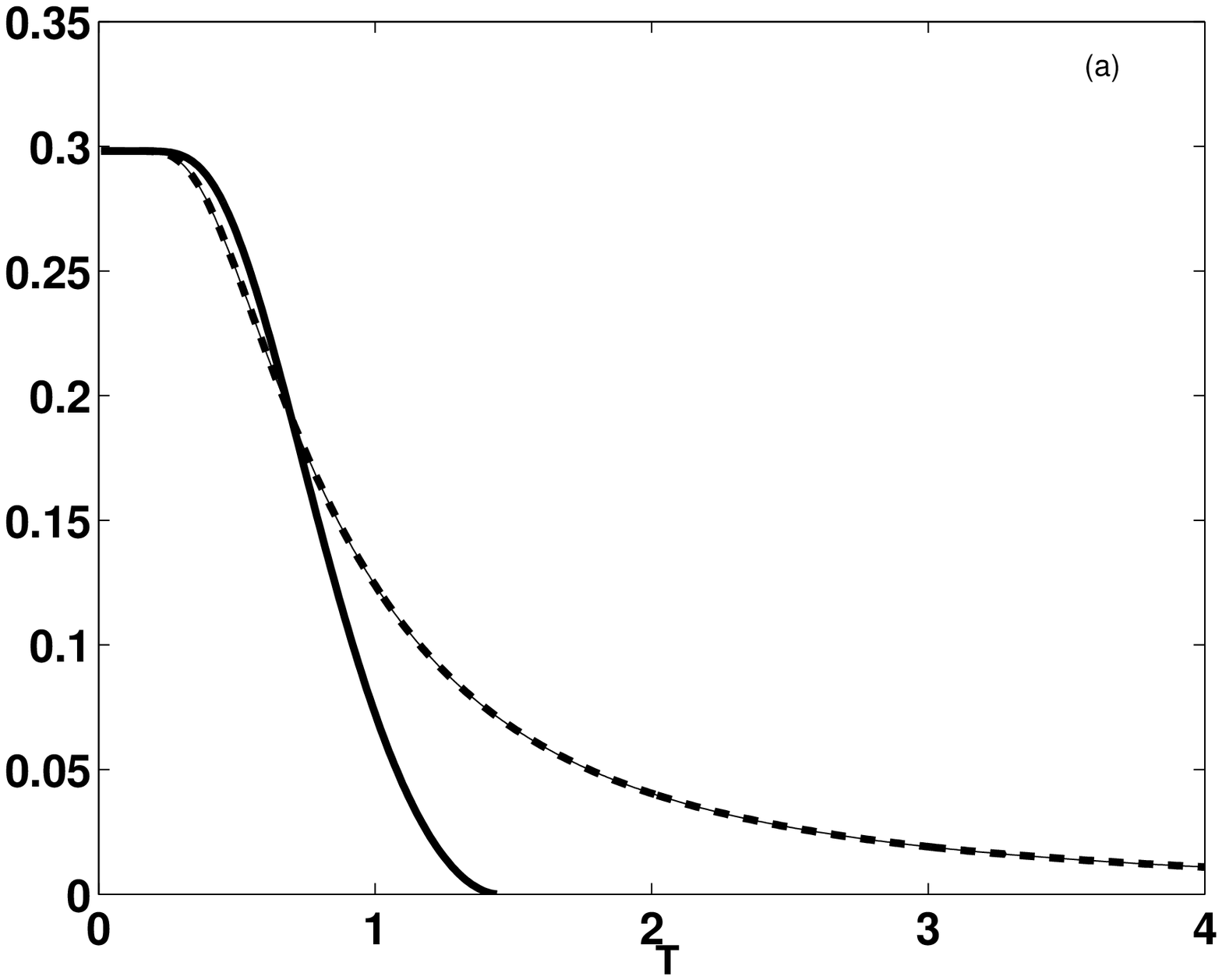}\\

\includegraphics[width=8cm,height=5cm]{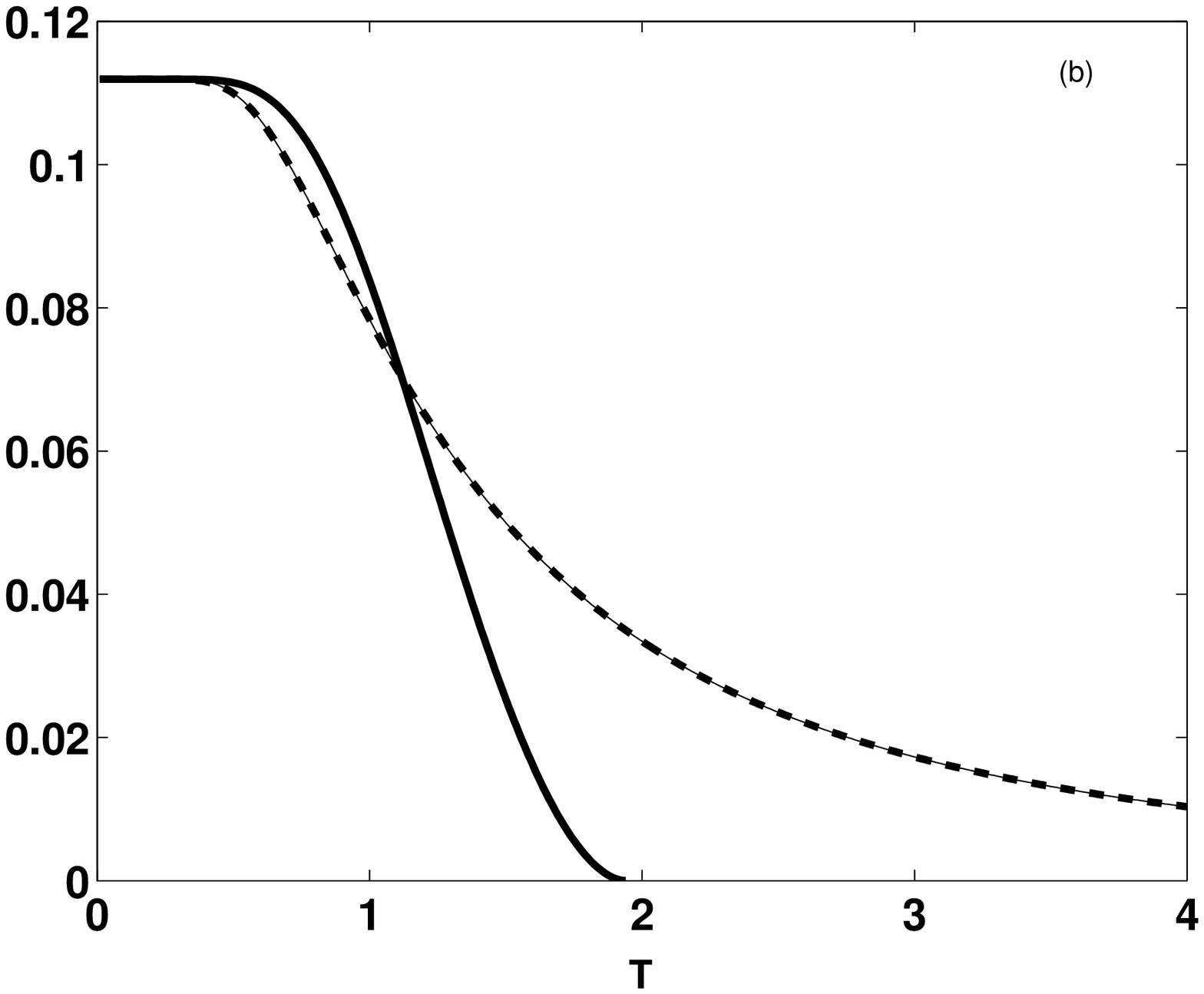}

Figure 4 : $QD$ and $CC$ (dashed line) and $EN$ (solid line) as a function  of the absolute temperature $T$ for (a) $B_1=-B_2=1$
(b) $B_1=-B_2=2$
\end{center}
\end{figure}

  We now deal with the case $B_2=-a B_1$, $a\neq 1$ and positive. $B_1=0$ satisfies both, $B_2=-aB_1$ and $B_2=-B_1$, so that $QD=CC$ at $B_1=0$, for all temperatures T. For a fixed temperature T, it turns out that $QD > CC$ for $B_1 \neq 0$ if $a>1$ and $CC > QD$ for $B_1 \neq0$ if $0 < a <1$. This is depicted in Fig. 5, for $a=2$ and $a=1/2$ for $T=1.5$. The dominance of $QD$ over $CC$ (or vice versa) varies continuously with $a$. This observation gives us the key to control the contributions of $QD$ and $CC$ to a two qubit thermal state in Heisenberg model via the continuous variation of the applied magnetic field.
  The behavior of concurrence in this case can be analyzed in a way similar to the case $B_2=-B_1, (a=1)$. From Fig. 5, we see that for the same temperature, the range over which concurrence vanishes depends on $a$, this range decreases monotonically with $a$.  Also, the peak position of concurrence (or, $EN$) on the $B_1$ axis shifts monotonically towards $B_1=0$ as $a$ increases. Thus the main entanglement features can be controlled by varying external magnetic fields.

   \begin{figure}[!ht]
\begin{center}
\includegraphics[width=8cm,height=5cm]{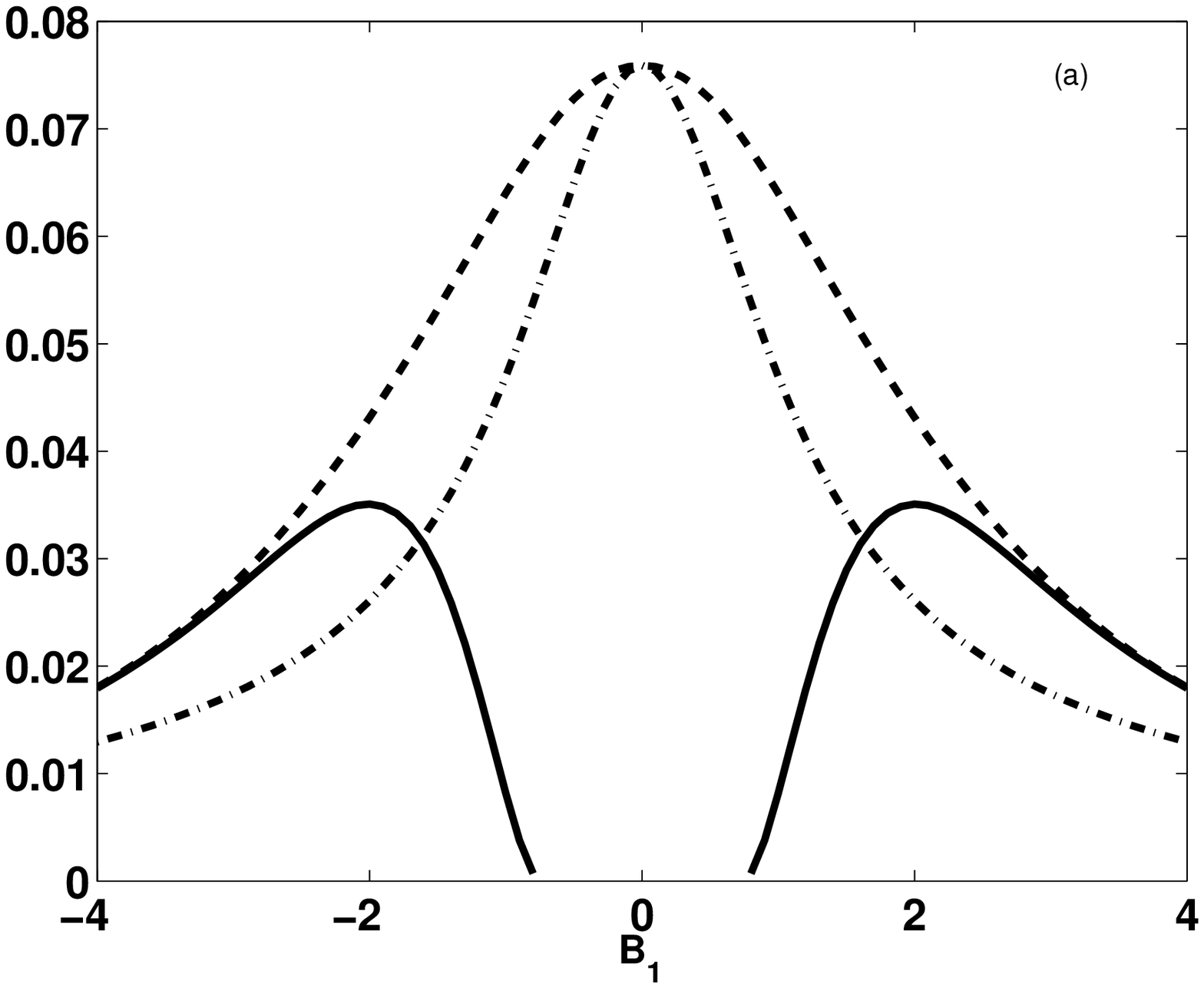}

\includegraphics[width=8cm,height=5cm]{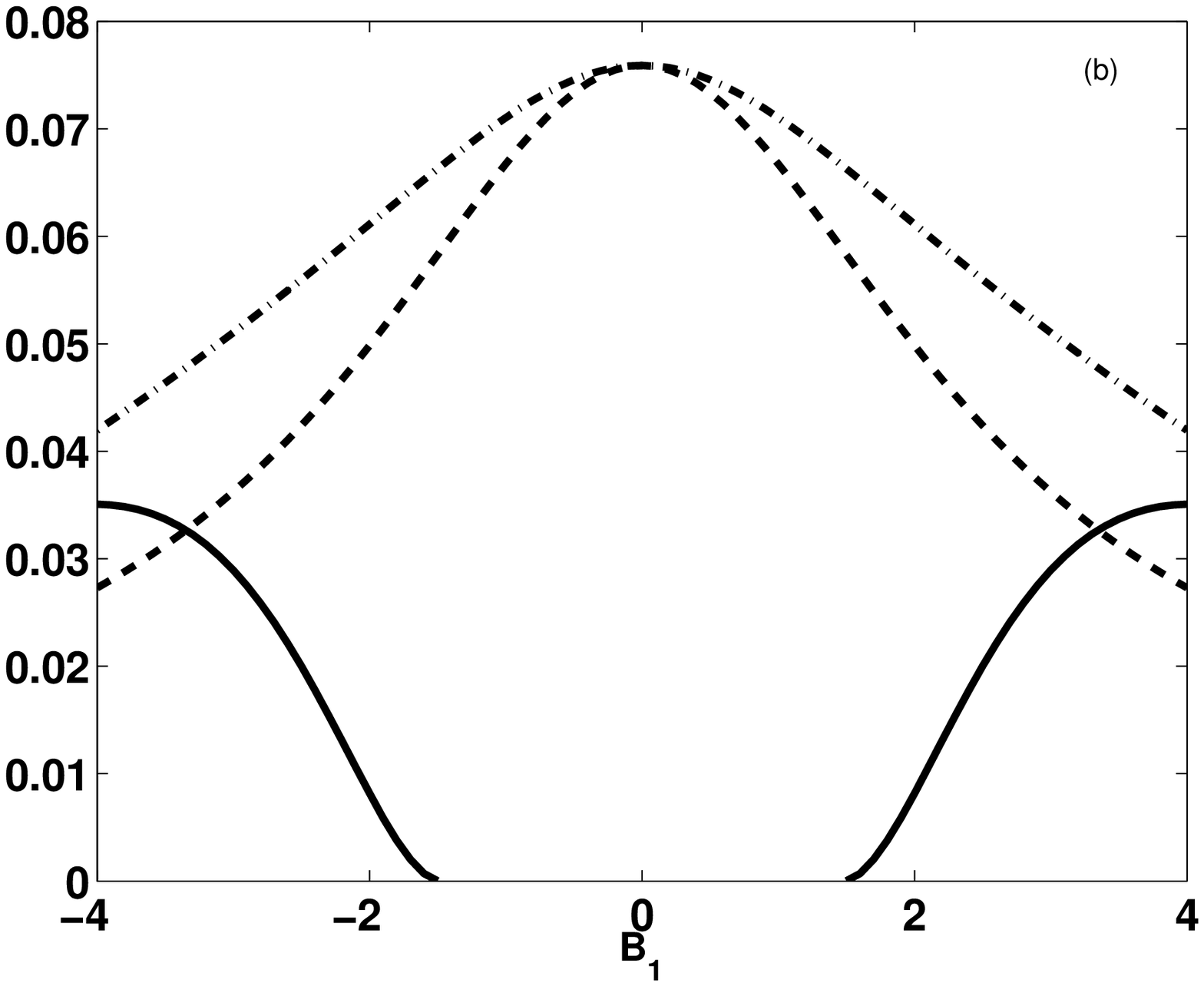}

Figure 5 : $QD$ (dashed line), $CC$ (dash-dotted line) and $EN$ (solid line) as a function  of external magnetic field $B_1$ at $T=1.5$  (a) $B_2=-2B_1$
(b) $B_2=-B_1/2$.
\end{center}
\end{figure}

Case II : $B_1=B_2$.

 For uniform external magnetic field $B_1=B_2$, Figs. 6, 7, 8, show the variation of $QD, CC$ and $EN$ with $B_1$ for temperatures $T=0.2, 0.9, 1.5$, respectively. We see that all three quantities are symmetric about $B_1=0$ where they have their maxima. Further, $QD > CC$, except at $B_1=0$, where $QD=CC$. For higher temperatures, the qualitative behavior of $QD$ and $CC$ remains the same, while $EN$ curve drops down below those of $QD$ and $CC.$ This can be qualitatively understood by looking at the thermal state given by
$$\rho=\frac{1}{Z}\big{[}e^{2B_1/{T}}|00\rangle\langle00|+e^{-2B_1/{T}}|11\rangle\langle11|$$
$$+e^{J/{T}}|\psi^-\rangle\langle\psi^-|+e^{-J/{T}}|\psi^+\rangle\langle\psi^+|\big{]}.\eqno{(14)}$$

For small temperatures, the entanglement of the thermal state is largely dictated by that of $|\psi^-\rangle$ and becomes dominant. At higher temperatures, admixture due to other states reduces the entanglement, so that $QD$ and $CC$ dominate. Such a complementary behavior of entanglement and discord can serve as a pointer towards a possible connection between them.

\begin{figure}[!ht]
\begin{center}
\includegraphics[width=8cm,height=5cm]{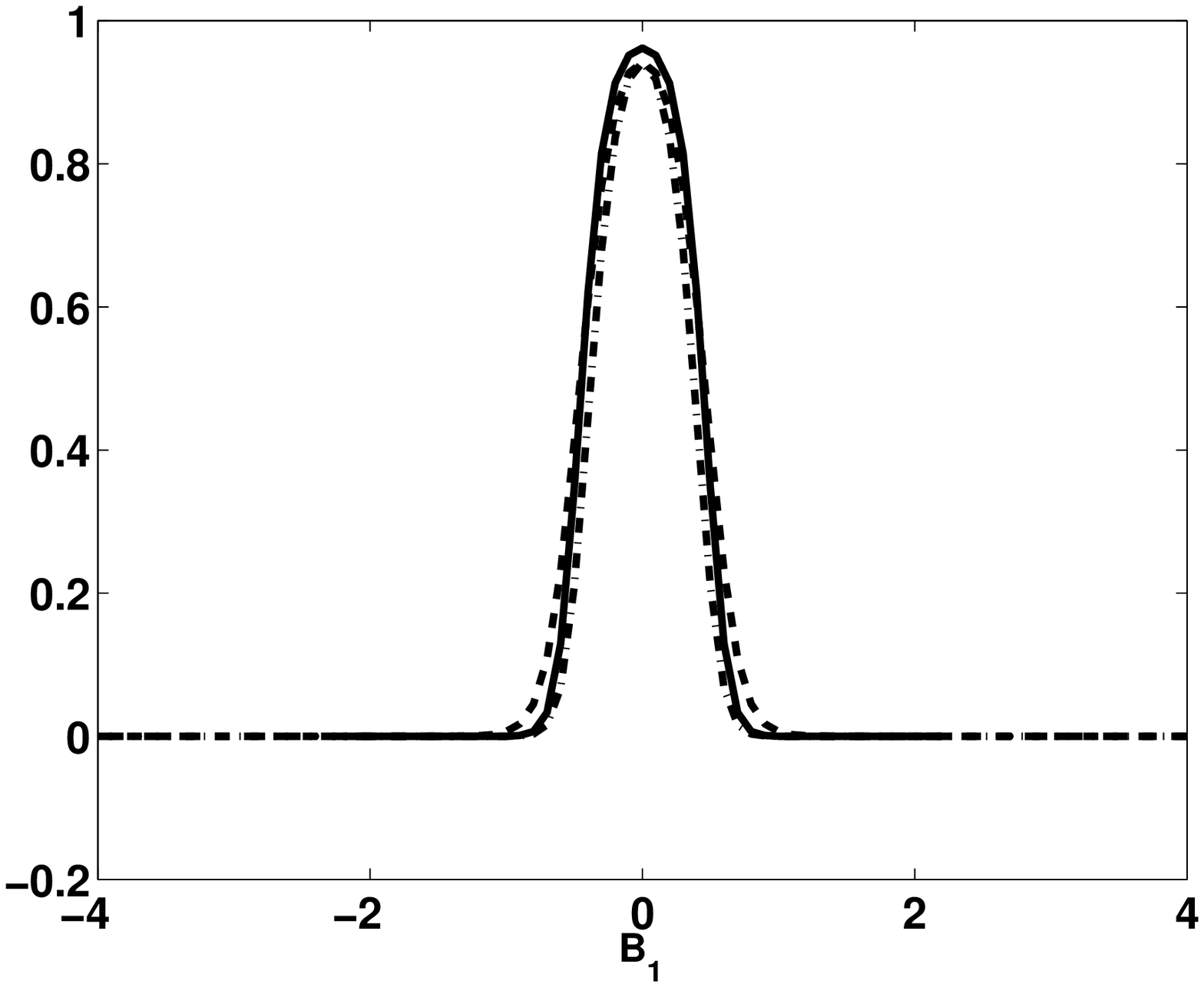}

Figure 6 : $QD$ (dashed line), $CC$ (dash-dotted line) and $EN$ (solid line) as a function  of external magnetic field $B_1$ where $B_1=B_2$ at $T=0.2$
\end{center}
\end{figure}

\begin{figure}[!ht]
\begin{center}
\includegraphics[width=8cm,height=5cm]{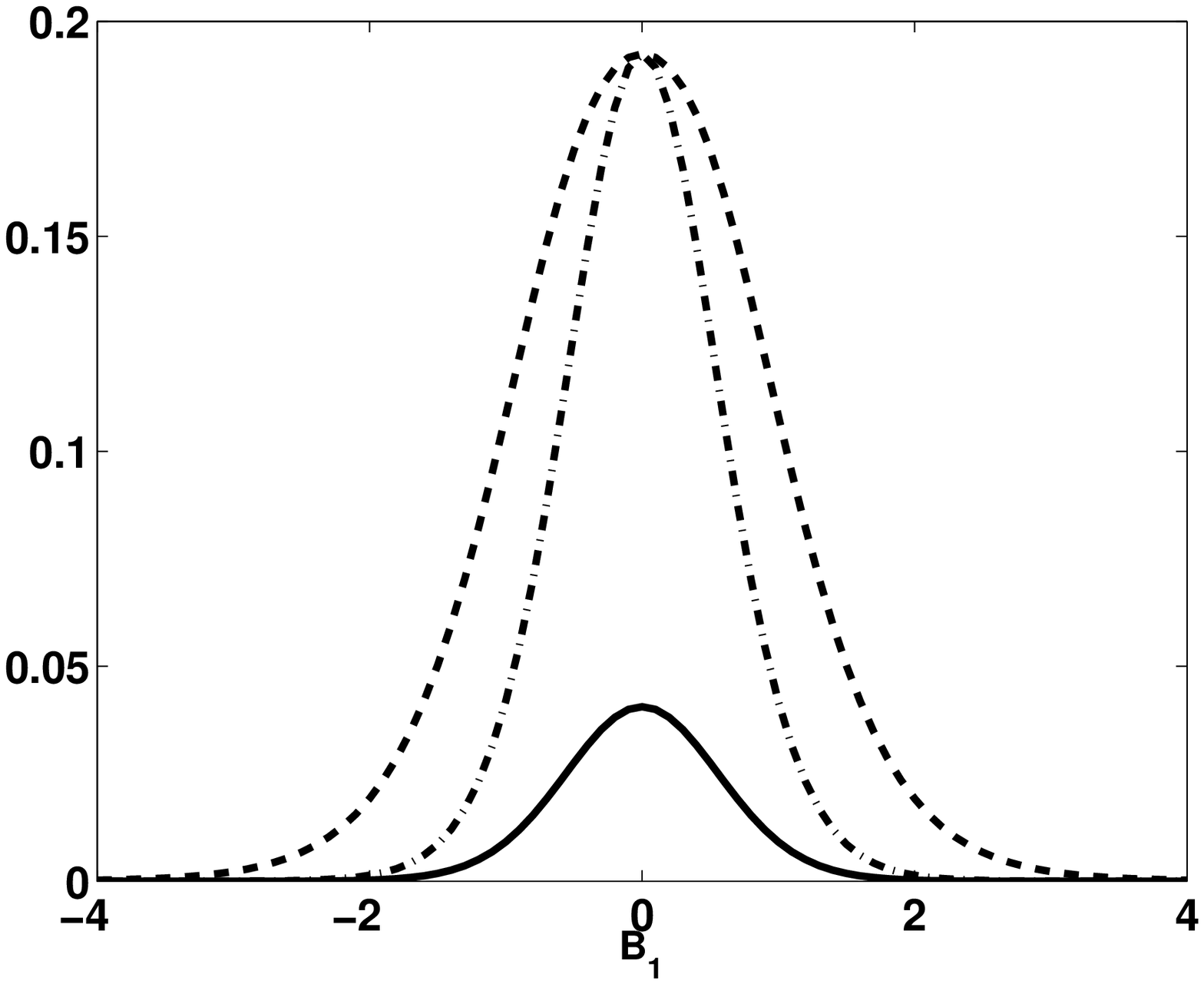}

Figure 7: $QD$ (dashed line), $CC$ (dash-dotted line) and $EN$ (solid line) as a function  of external magnetic field $B_1$ where $B_1=B_2$ at $T=0.9$
\end{center}
\end{figure}

\begin{figure}[!ht]
\begin{center}
\includegraphics[width=8cm,height=5cm]{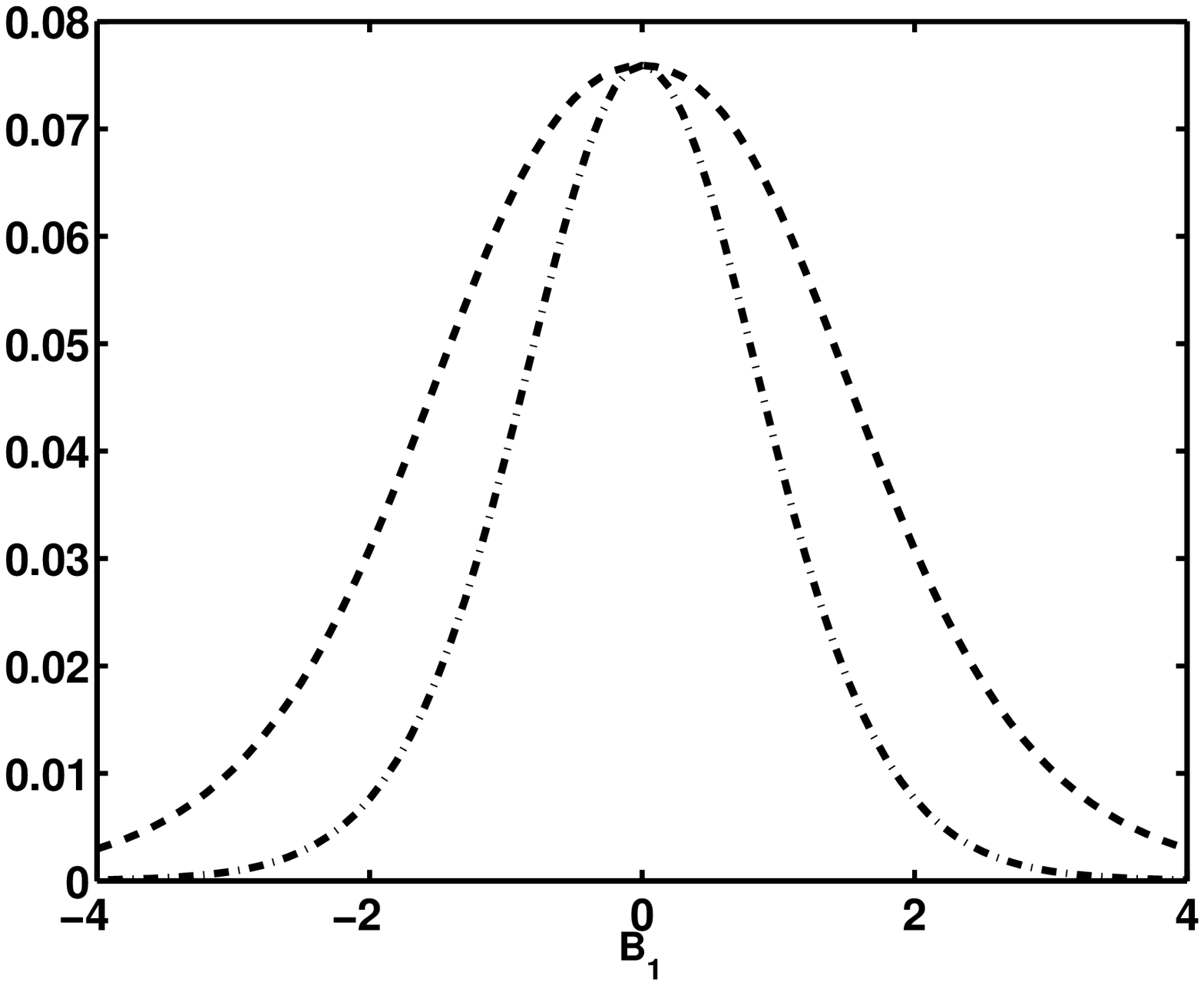}

Figure 8 : $QD$ (dashed line), $CC$ (dash-dotted line) and $EN$ (solid line) as a function  of external magnetic field $B_1$ where $B_1=B_2$ at $T=1.5$
\end{center}
\end{figure}

Fig. 9 shows the variation of $QD, CC$ and $EN$ with temperature for $B_1=B_2=1$ and $B_1=B_2=2$. We see that for high temperatures, $QD$ hugely dominates $EN$, showing the robustness of $QD$ with temperature. As temperature becomes large $QD$ and $CC$ converge towards each other. For larger values of $B_1$ this happens at higher temperatures. As the temperature increases, all the coefficients in the thermal mixture Eq.(14) tend to be equal and the thermal state approaches random mixture. Thus it seems that $QD$ and $CC$ approach each other as an arbitrary thermal state approaches a random mixture. Obviously, for random mixture $\rho=\frac{1}{4}(I\otimes I), ~ QD=CC=0.$ A quantitative analysis of the relative behaviors of $QD$ and $CC$ with temperature will be very interesting, but possibly have to wait for further developments in the theory.\\

\begin{figure}[!ht]
\begin{center}
\includegraphics[width=8cm,height=5cm]{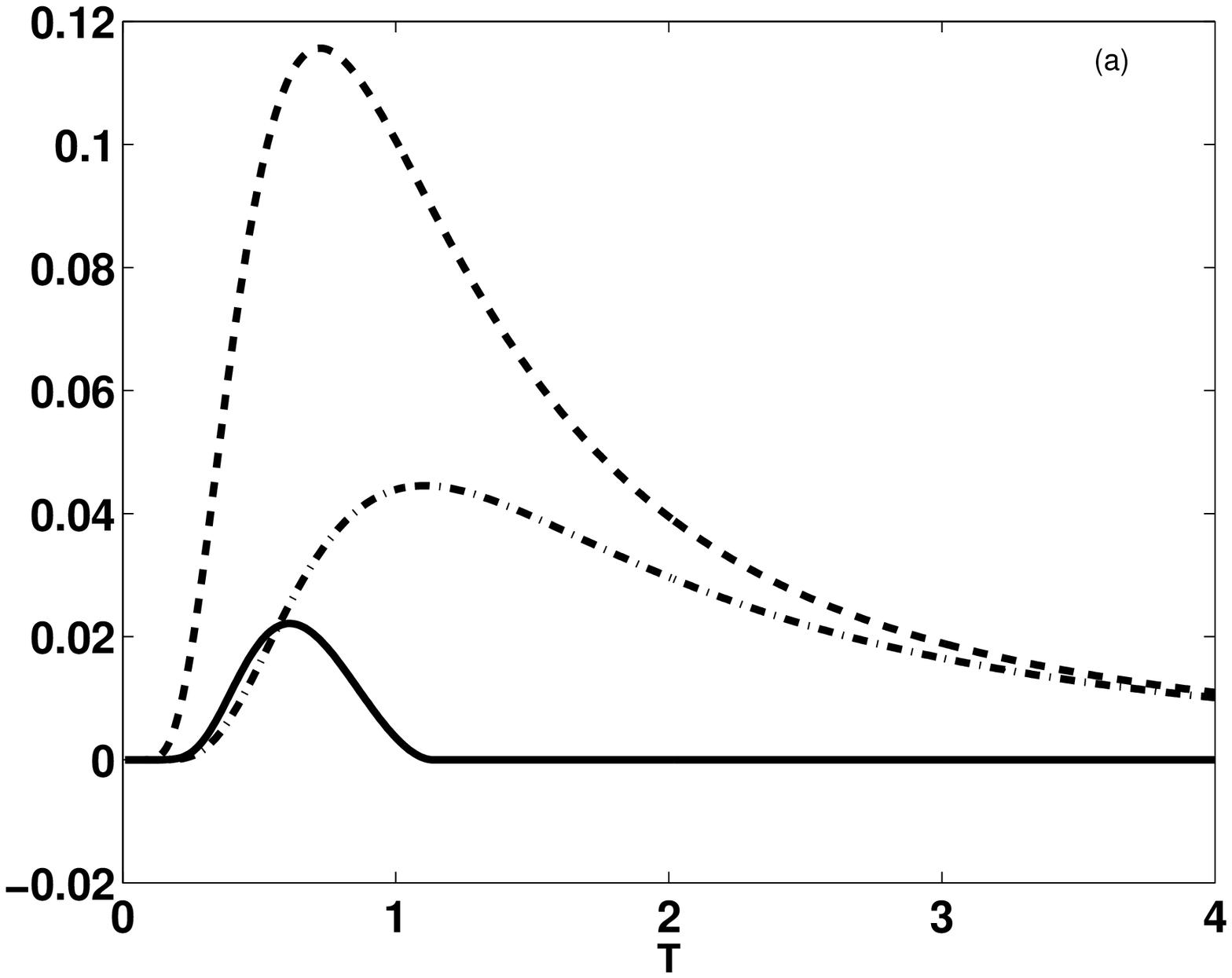}

\includegraphics[width=8cm,height=5cm]{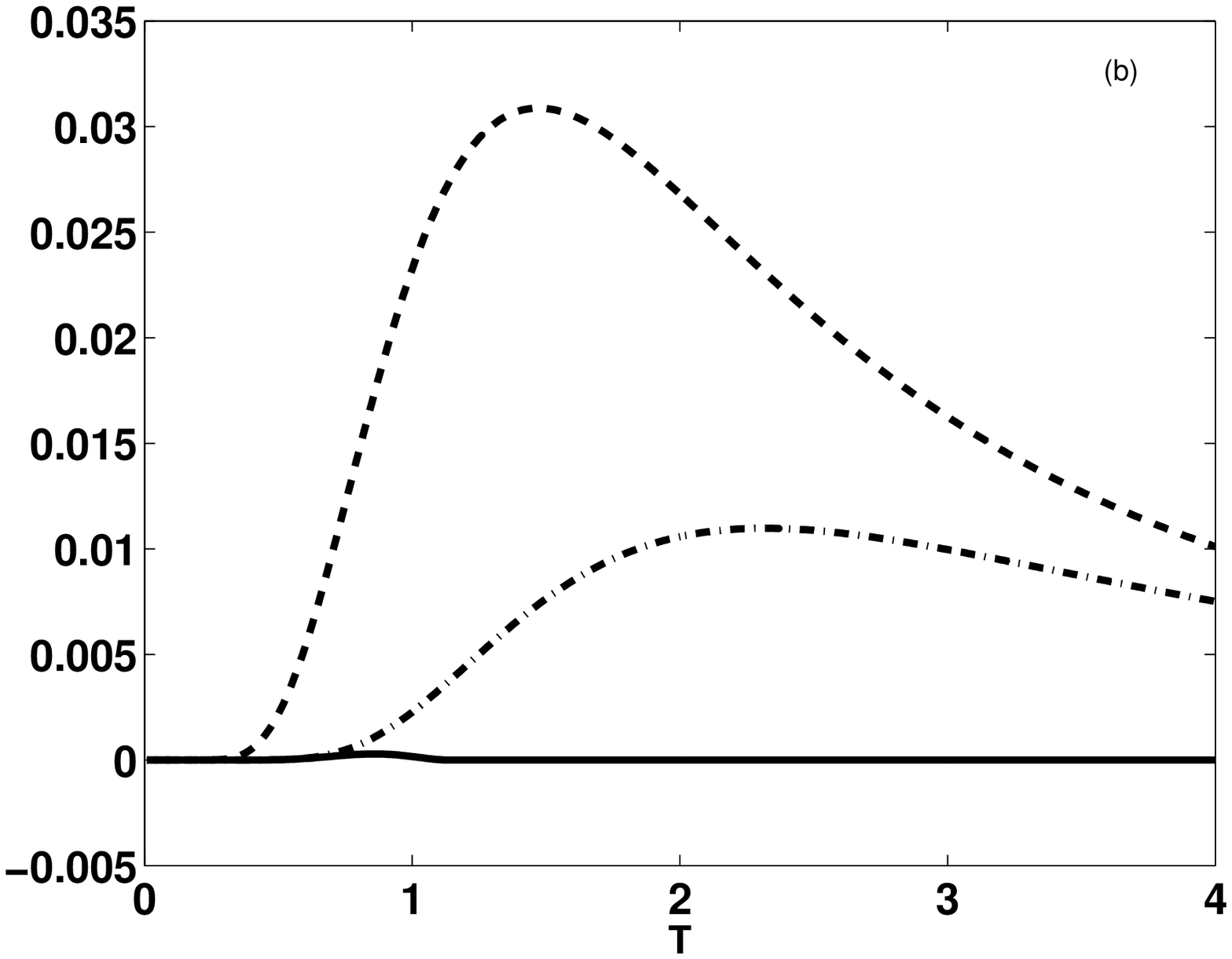}

Figure 9 : $QD$ (dashed line), $CC$ (dash-dotted line) and $EN$ (solid line) as a function  of the absolute temperature $T$ for (a) $B_1=B_2=1$
(b) $B_1=B_2=2$
\end{center}
\end{figure}

\begin{figure}[!ht]
\begin{center}
\includegraphics[width=8cm,height=5cm]{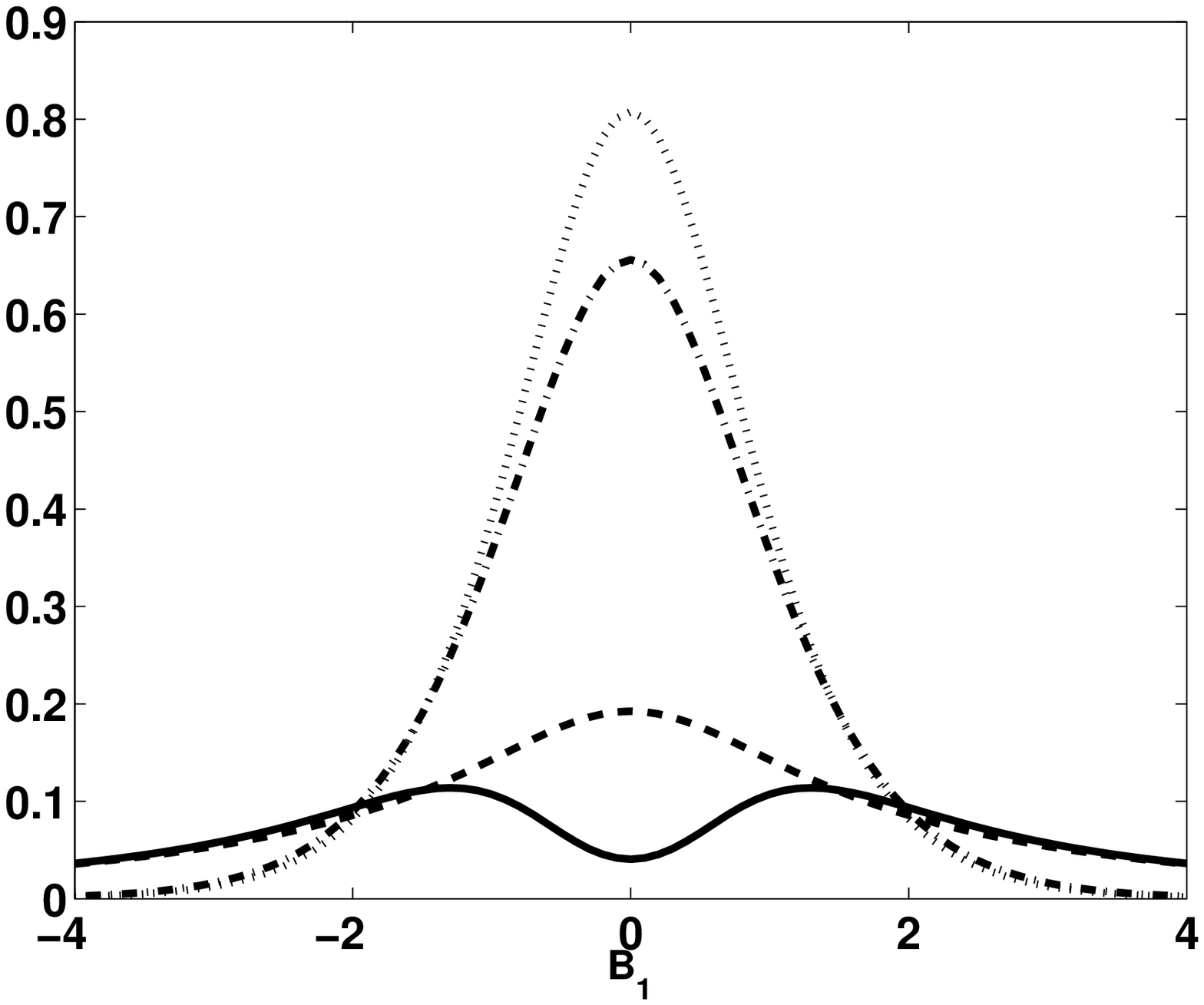}

Figure 10 : $EN_{AB}$ (solid line), $QD_{\overleftarrow{AB}}$  (dashed line), $EN_{AE}$  (dot line) and $QD_{\overleftarrow{AE}}$  (dash-dotted line), as a function  of external magnetic field $B_1=-B_2$ at $T=0.9$
\end{center}
\end{figure}

It will be intresting to connect our results with the monogomy relations between the $EN$ and the classical correlation \cite{koa} of two subsystems (qubits) and the environment

$$EN_{AB}+CC_{\overleftarrow{AE}}=S_A, $$

$$EN_{AE}+CC_{\overleftarrow{AB}}=S_A, \eqno{(15)}$$
and the relation between $EN$ and $QD$ \cite{fanc},
$$EN_{AB}+EN_{AE}=QD_{\overleftarrow{AB}}+QD_{\overleftarrow{AE}},\eqno{(16)}$$
showing us that $EN$ and $QD$ always exist is pairs. Here $A$,$B$ label the qubits and $E$ stands for the environment. We assume that environment (heat bath) comprises the universe  minus the qubits $A$ and $B$ so that the state $\rho_{ABE}$ is a pure state. Since the variation of all the quantities pertaining to the system $AB$ with $B_1$ and $T$ are obtained form the $XX$ model, we can use Eq.(15,16) to find the corresponding dependence of $EN_{AE}$ and $QD_{\overleftarrow{AE}}$ on $B_1$ and $T$. Figs. 10 and 11, (for $B_1=-B_2$) show the variation of $EN_{AE}$ and $QD_{\overleftarrow{AE}}$ with $B_1$ and $T$.\\

The monogamic relations also help us establish a necessary and sufficient condition for $QD_{\overleftarrow{AB}}=CC_{\overleftarrow{AB}}$ when the environment is present.
This is : $QD_{\overleftarrow{AB}}=CC_{\overleftarrow{AB}}$ if and only if $\frac{1}{2}I_{AB}=EN_{AE}+EN_{AB}-QD_{\overleftarrow{AE}}.$ To prove the necessity we note that when $QD_{\overleftarrow{AB}}=CC_{\overleftarrow{AB}},$ (that is, $QD_{\overleftarrow{AB}}=\frac{1}{2}I_{AB}$), Eq.(16) can be written as $$\frac{1}{2}I_{AB}=EN_{AE}+EN_{AB}-QD_{\overleftarrow{AE}}.\eqno{(17)}$$ Now suppose Eq.(17) is true. Then using Eq.(16) we have $$QD_{\overleftarrow{AB}}=\frac{1}{2}I_{AB}$$ which implies $QD_{\overleftarrow{AB}}=CC_{\overleftarrow{AB}}.$ Figs.12 and 13, show the variation of both sides of Eq.(17) with $B_1$ and $T$ which establishes Eq.(17) for the XX model.

\begin{figure}[!ht]
\begin{center}
\includegraphics[width=8cm,height=5cm]{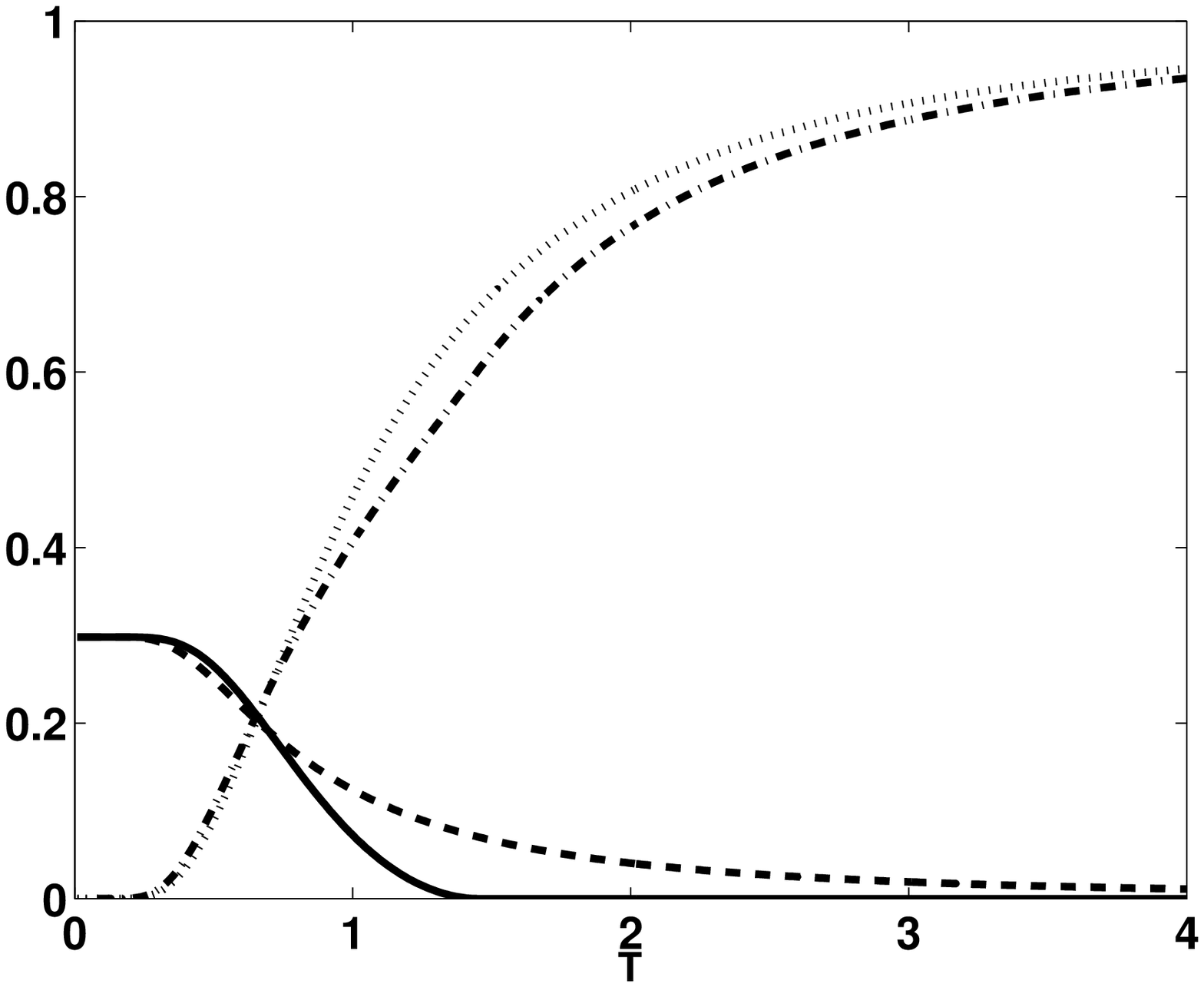}

Figure 11 : $EN_{AB}$ (solid line), $QD_{\overleftarrow{AB}}$  (dashed line), $EN_{AE}$  (dot line) and $QD_{\overleftarrow{AE}}$  (dash-dotted line), as a function  of the absolute temperature $T$ for  $B_1=-B_2=1.$

\end{center}
\end{figure}

\begin{figure}[!ht]
\begin{center}

\includegraphics[width=8cm,height=5cm]{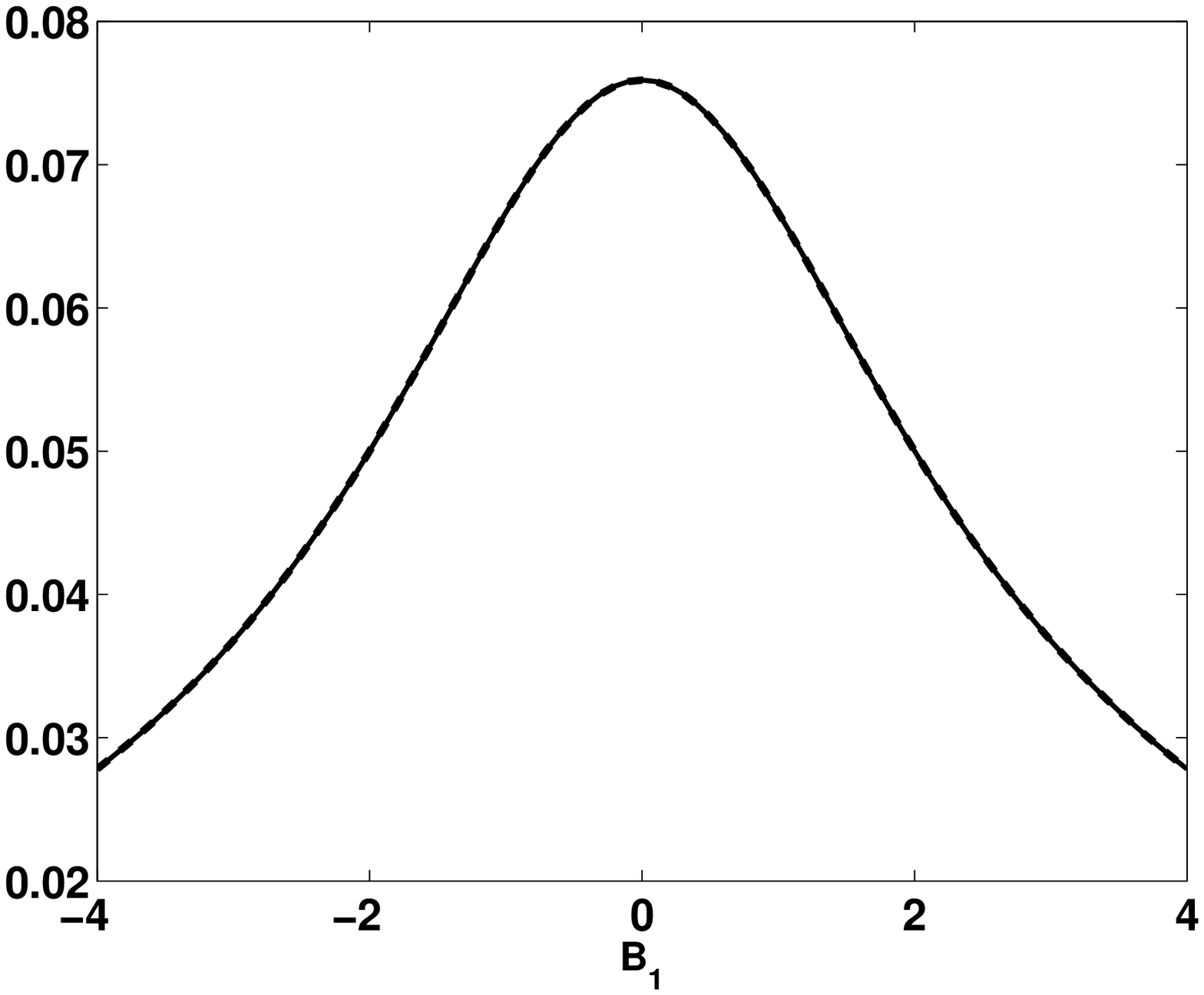}

  Figure 12 : $\frac{1}{2}I_{AB}$ (solid line), $EN_{AE}+EN_{AB}-QD_{\overleftarrow{AE}}$  (dashed line), as a function of external magnetic field $B_2=-B_1$ at $T=1.5$

\end{center}
\end{figure}

\begin{figure}[!ht]
\begin{center}

\includegraphics[width=8cm,height=5cm]{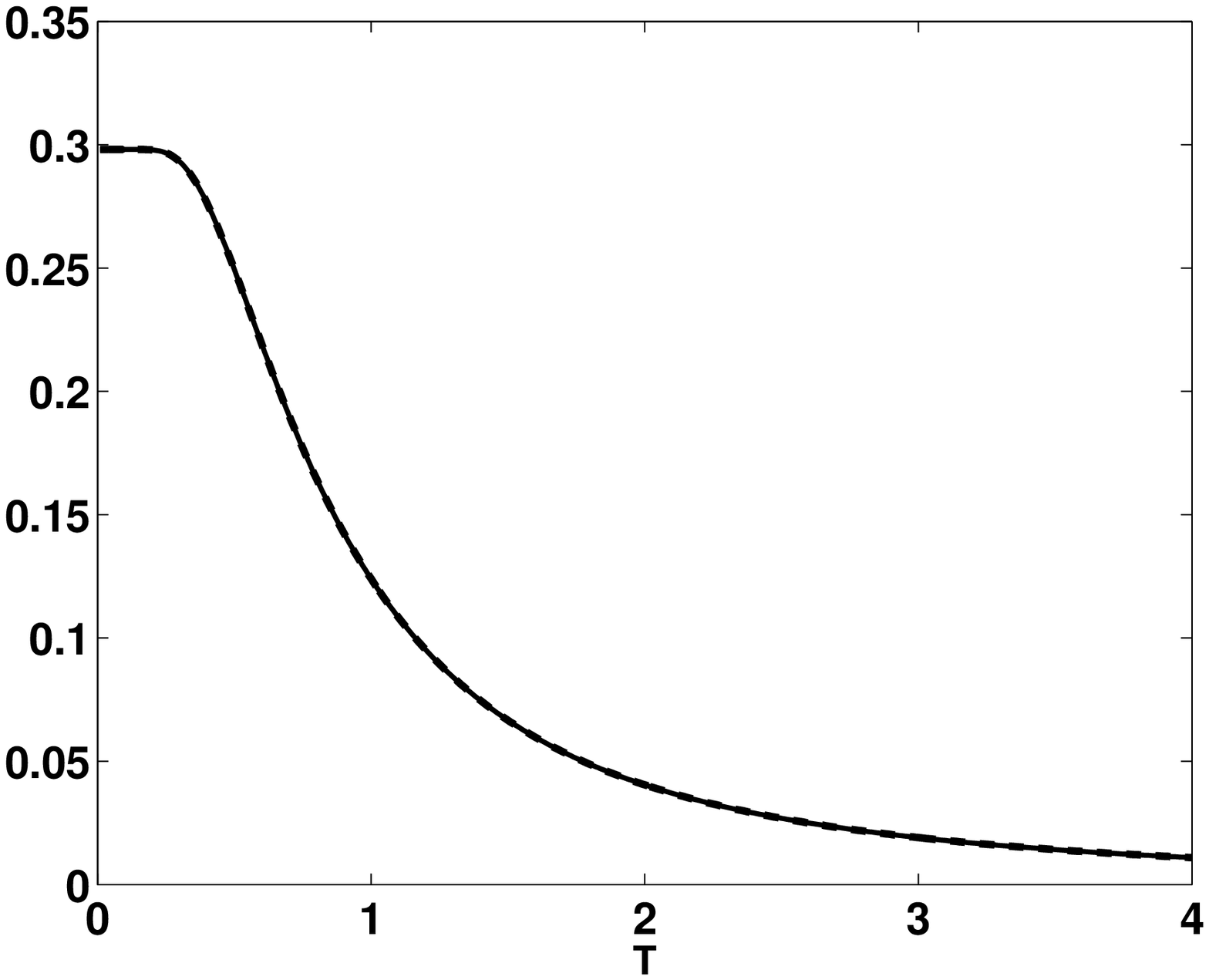}

Figure 13 : $\frac{1}{2}I_{AB}$ (solid line), $EN_{AE}+EN_{AB}-QD_{\overleftarrow{AE}}$  (dashed line), as a function  of the temperature $T$ for $B_2=-B_1=-1.$

\end{center}
\end{figure}

\emph{Summary.}
In this paper we have studied the variation of $QD, CC$ and $EN$ in two qubit XX Heisenberg chain as functions of independently varied magnetic fields $B_1$ and $B_2$ on each qubit and also with temperature. We deal with two cases $B_1= - B_2$ (nonuniform field) and $B_1=B_2$ (uniform field). Our first observation is the complementary behavior of entanglement and $QD/CC$. For the nonuniform magnetic field, we get the interesting observation that $QD$ and $CC$ are equal for all $B_1= - B_2$ values as well as all temperatures. Surprisingly, this observation is explained quite simply, using the symmetric form of the thermal state. A very interesting observation is that the relative contributions of $QD$ and $CC$ can be tunably controlled by varying the applied magnetic field. Another interesting finding is that the equality of $QD$ and $CC$ of the subsystem (qubits) imposes a constraint on the distribution of $QD$ and $EN$ over the subsystem and its environment. Further investigation of general Heisenberg models like XXZ along these lines may turn out to be interesting and fruitful.\\

\emph{Appendix.}  We prove the following statement.

  If  the quantum state has  the Bloch representation \cite{hj}
$$\rho=\frac{1}{4}[I\otimes I+\sum_{i=1}^3 c_i \sigma_i\otimes \sigma_i], \eqno{(A1)}$$
and $c_i=c_j>c_k$ and $c_k=-c_i^2$ where $i\neq j \neq k \in \{1,2,3\}$, then this state contains the same amount of quantum and classical correlation $(QD=CC)$.

\emph{Proof}: In Ref. \cite{luo}  S. Luo evaluated analytically the quantum discord  for a large family of two-qubit states, which have the maximally mixed marginal and their Bloch representation is

  $$\rho=\frac{1}{4}[I\otimes I+\sum_{i=1}^3 c_i \sigma_i\otimes \sigma_i].$$

  For this class of quantum states the quantum mutual information is\\
  $ \mathcal{I}(\rho)=\frac{1}{4}[(1-c_1-c_2-c_3) log_2(1-c_1-c_2-c_3)+(1-c_1+c_2+c_3) log_2(1-c_1+c_2+c_3)+(1+c_1-c_2+c_3) log_2(1+c_1-c_2+c_3)+(1+c_1+c_2-c_3) log_2(1+c_1+c_2-c_3)].$

   We substitute the conditions above in the quantum mutual information. Puting $c=c_1=c_2 >c_3$ and $c_3=-c^2$, we get,\\
  $ \mathcal{I}(\rho)=\frac{1}{4}[(1-2c+c^2) log_2(1-2c+c^2)+(1-c^2) log_2(1-c^2)+(1-c^2) log_2(1-c^2)+(1+2c+c^2) log_2(1+2c+c^2)]$

   After some algebraic simplification, we get\\
  $ \mathcal{I}(\rho)=(1-c) log_2(1-c)+(1+c) log_2(1+c)$\\
  which equals $2 CC$ as in Ref. \cite{luo}.
  It is also easy to check that the above argument goes through when  $c=c_1=c_3 >c_2$ and $c_2=-c^2$ and when $c=c_3=c_2 >c_1$ and $c_1=-c^2$,
to get $ \mathcal{I}(\rho)=2CC$.
Thus, $$QD(\rho)=\mathcal{I}(\rho)-CC(\rho)=CC(\rho).$$

\emph{Acknowledgments:}\\
ASMH thanks Pune University for hospitality during his visit when this work was carried out. PSJ thanks BCUD Pune University for financial support. We thank Guruprasad Kar and Prof. R. Simon for encouragement. We thank the anonymous referees whose comments have contributed towards the improvement of this paper.

\end{document}